\documentclass[english,aps,pre,twocolumn,showpacs,floatfix]{revtex4-1}
\usepackage[T1]{fontenc}
\usepackage[latin1]{inputenc}
\usepackage{amsmath}
\usepackage{graphicx}
\usepackage{amssymb}
\usepackage{dcolumn} 
\usepackage{bm} 
\usepackage{cleveref}

\usepackage{xcolor}
\usepackage{framed}
\colorlet{shadecolor}{blue!15}
\usepackage[none]{hyphenat}

\makeatletter

\usepackage{amsfonts}
\usepackage{epsfig}
\usepackage{dsfont}
\usepackage{babel}

\makeatother

\RequirePackage[normalem]{ulem} 
\RequirePackage{color}\definecolor{RED}{rgb}{1,0,0}\definecolor{BLUE}{rgb}{0,0,1} 

\begin{document}

\title{Patterns for the waiting time in the context of discrete-time stochastic processes}

\author{Tayeb \surname {Jamali}}
\affiliation{Department of Physics, Shahid Beheshti University, G.C., Evin, Tehran 19839, Iran}

\author{G.R. \surname {Jafari}}
\affiliation{Department of Physics, Shahid Beheshti University, G.C., Evin, Tehran 19839, Iran}
\affiliation{Center for Network Science, Central European University, H-1051, Budapest, Hungary}

\author{S. \surname {Vasheghani Farahani}}
\affiliation{Department of Physics, Tafresh University, Tafresh 39518 79611, Iran}

\date{\today}

\begin{abstract}

The aim of this study is to extend the scope and applicability of the level-crossing method to discrete-time stochastic processes and generalize it to enable us to study multiple discrete-time stochastic processes. In previous versions of the level-crossing method, problems with it correspond to the fact that this method had been developed for analyzing a continuous-time process or at most a multiple continuous-time process in an individual manner. However, since all empirical processes are discrete in time, the already-established level-crossing method may not prove adequate for studying empirical processes. Beyond this, due to the fact that most empirical processes are coupled; their individual study could lead to vague results. To achieve the objectives of this study, we first find an analytical expression for the average frequency of crossing a level in a discrete-time process, giving the measure of the time experienced for two consecutive crossings named as the ``waiting time''. We then introduce the generalized level-crossing method by which the consideration of coupling between the components of a multiple process becomes possible. Finally, we provide an analytic solution when the components of a multiple stochastic process are independent Gaussian white noises. The comparison of the results obtained for coupled and uncoupled processes measures the strength and efficiency of the coupling, justifying our model and analysis. The advantage of the proposed method is its sensitivity to the slightest coupling and shortest correlation length.

\end{abstract}


\maketitle

\section{Introduction}

\subsection{Motivation }

The level-crossing method has a long history in studying stationary stochastic processes~\cite{kac1943,rice1944}. The matter of interest in the level crossing method is to extract the statistical properties of a stationary process by its upcrossings through a specific level. This is very important from a mathematical point of view, since it is closely related to the problem of extremes in random processes~\cite{LLR}. In the context of the present study, a more hands-on practical aspect of level crossing named the ``waiting time'' is of interest. In the level-crossing method, the average time between two consecutive crossings of the level is a measure of the ``waiting time''. The concept of waiting time has been studied in various fields, e.g., statistical physics \cite{SBM,CB}, material physics \cite{KPAKKS,IM,KVLM}, quantum physics \cite{RPG,DGFMI}, solar physics \cite{Aschwanden,TCLA}, fluid dynamics \cite{VLSL,VLFASL}, and finance \cite{JMFRM,DJSS,MSFDS}. 

It should be noted that prior to the present study the level-crossing method has been mostly implemented for studying \emph{continuous-time} processes, paying less attention on \emph{discrete-time} processes either for individual or coupled processes. Now two questions may arise. Firstly, why does studying level crossing for discrete-time processes matter? secondly, why are we required to study coupled processes? The significance of studying level crossing for discrete-time processes comes from the fact that all empirical processes are discrete in time. This importance has been amplified by the inconsistency observed between the level-crossing results for continuous- and discrete-time processes. Therefore, we need to study the level-crossing method for discrete-time processes before applying it to empirical data. To address the second question, we should note that the level-crossing method had been introduced for analyzing a single process. In other words, it works fine for the individual study of processes. However, we sometimes need to study \emph{multiple} stochastic processes, where some sort of coupling or coexistence between processes exist. This motivated us to generalize the level-crossing method for studying multiple discrete-time processes. The generalized level-crossing method can then be applied to measure the coupling between processes. 

Various techniques have been developed for studying the coupling between processes, e.g. the cross-correlation and cross-spectral density~\cite{RandomVibrations,TimeSeries}, random matrix theory \cite{lcbp,pgras,pgras2,pgrags,mnsv}, the detrended cross-correlation analysis (DXA)~\cite{ps,djkms,sj}, partial-DXA~\cite{partialDXA}, coupled-DXA~\cite{coupledDXA}, and the detrending moving-average cross-correlation analysis~\cite{JZ}. These methods miss the contribution of local effects due to the fact that the act of averaging constitutes their back bone, bringing up the idea of using a method independent of averaging which only counts on local effects. This is where the quest for the generalized level crossing is intensified. In the present study we develop the level-crossing method to account for multiple stochastic processes. We find an analytic result for a multiple process whose components are independent Gaussian white noises. This result plays the role of a criterion of a fully uncorrelated multiple process. By comparing the analytic result with that of a coupled multiple discrete-time process, the efficiency of the coupling in a multiple process is obtained.

\subsection{Background}

Level crossing has been mainly implemented when dealing with continuous-time stationary stochastic processes, especially the famous class of Gaussian processes. In this subsection, we glance over the level-crossing method for a continuous-time stationary process $X=\{X_t, t\ge 0\}$. An upcrossing of the level $x$ at time $t_0$ is bound to occur if we have $X_t < x$ in $(t_0-\delta,t_0)$ and $X_t > x$ in $(t_0,t_0+\delta)$, where $\delta$ is the neighborhood radius. The number of upcrossings of the level $x$ by $X$ over the interval $[0,t]$ is denoted by $N_t^{+}(x)$. As $N_t^{+}(x)$ is itself a random variable, the mean number of upcrossings shown by $\langle N_t^{+}(x) \rangle$ is the quantity of interest. The general form of $\langle N_t^{+}(x) \rangle$ for the continuous-time stationary process $X$ is proved to be given by~\cite{LLR}
\begin{equation}
\label{eq: LC for stationary SP}
\langle N_t^{+}(x) \rangle =t \int_0^\infty dz\; z\,p(x,z),
\end{equation}
where $p(x,z)$ is the joint probability density of $(X,\dot{X})$ containing all information about the level-crossing characteristics of a stationary process $X$. Note that the parameter $t$ on the right-hand side of this equality comes from the stationarity of $X$. 

We turn our attention now to the case of stationary Gaussian processes. For a standardized stationary Gaussian process with correlation function $r(\tau) = \langle X_t X_{t+\tau} \rangle$~\cite{comment1}, the mean number of upcrossings of $X$ is given by~\cite{LLR}
\begin{equation}
\label{eq: LC for stationary Gaussian SP}
\langle N_t^{+}(x) \rangle =\frac{t}{2\pi} \lambda_2^{1/2} \exp \left( -\frac{x^2}{2} \right),
\end{equation}
where $\lambda_2$ is the second moment of the spectral density function $f(\lambda) = \int_{-\infty}^{\infty} r(\tau) e^{-i\lambda \tau}\,d\tau/2\pi$. The Gaussian white noise is a special case in the family of stationary Gaussian processes whose correlation function is given by the Dirac delta function, $\delta(\tau)$. It should be noted that for a Gaussian white noise, Eq.~(\ref{eq: LC for stationary Gaussian SP}) diverges because its second spectral moment $\lambda_2$ is infinity. 

Numerical studies have shown that Eq.~(\ref{eq: LC for stationary Gaussian SP}), which is for continuous-time stationary Gaussian processes, is inconsistent with the level-crossing results of discrete-time stationary Gaussian processes. Therefore, studying the level-crossing method for discrete-time processes could help avoid this discrepancy. The two following sections are devoted to this issue. In the fourth section, we create a criterion for measuring the coupling in multiple processes. Note that in this paper since we only deal with discrete-time processes, for simplicity the term ``stochastic process'' is used instead of the term ``discrete-time stochastic process'', unless for emphasizing the type of process.

\section{Level crossing for discrete-time stochastic processes}

Consider a discrete-time stochastic process represented by $\{X\}\equiv\{X(t_1), X(t_2),\dots,X(t_n)\}$. As stated earlier, the intention is to find the number of upcrossings, $N_n^+(x)$, of a typical level, $x$, by the process $\{X\}$ in the time period between $t_1$ and $t_n$. Mathematically speaking, an upcrossing of the level $x$ at time $t_i$ is when we have $X(t_{i-1})< x$ and $X(t_{i+1})> x$. Since $\{X\}$ is a stochastic process, $N_n^+(x)$ would be a random variable. Therefore, its ensemble average which is denoted by $\langle N_n^+(x)\rangle$ is the quantity of interest. In case of a stationary stochastic process, $\langle N_n^+(x)\rangle$ would become proportional to $n$ with a proportionality constant represented by $\nu_x^+$. Note that $\nu_x^+$ is the average frequency of the upcrossings of the level $x$, where its inverse $\tau_{x}=1/\nu_{x}^+$ gives the waiting time expressed as the average time expected for two consecutive upcrossings by $\{X\}$. The average frequency $\nu_x^+$ for a stationary process $\{X\}$ is given by
\begin{equation}
\label{eq: 1d nu}
\nu_x^+ = \int_{-\infty}^{x} dx_1 \int_{x}^{\infty} dx_2 \; p_X(x_1,x_2),
\end{equation}
where $p_X(x_1,x_2)$ is the two-points joint probability density of the process $\{X\}$ for two successive points $x_1$ and $x_2$. The right-hand side of  Eq.~(\ref{eq: 1d nu}) is nothing but the occurrence probability of an upcrossing of the level $x$ by two successive points of $\{X\}$.

\begin{figure}[!t]
\centering
\includegraphics[width=1\linewidth]{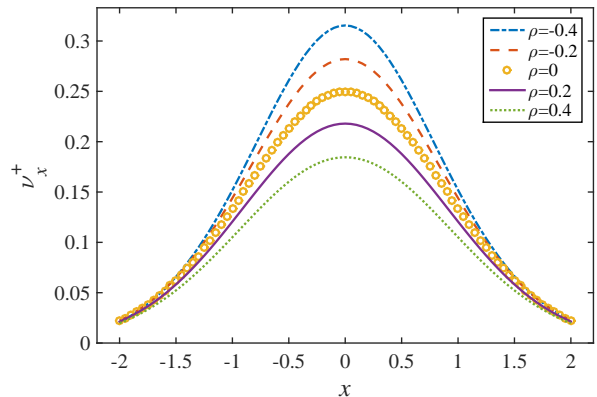}
\caption{The average frequency of the upcrossings of the level $x$ for various values of correlation $\rho$ in a discrete-time stationary Gaussian process.}
\label{fig: one-dimensional level-crossing frequency}
\end{figure}

For a standardized stationary Gaussian process, we have
\begin{equation}
\label{eq: two-points joint pdf}
p_X(x_1,x_2) = \frac{1}{2\pi \sqrt{1-\rho^2}} \exp{\left\{-\frac{x_1^2-2\rho x_1x_2+x_2^2}{2(1-\rho^2)}\right\}},
\end{equation}
where $\rho = \langle x_1 x_2\rangle $ is the correlation between two successive points $x_1$ and $x_2$. Substituting this equation into Eq.~(\ref{eq: 1d nu}) gives
\begin{align}
\label{eq: 1d nu for Gaussian processes}
\nonumber
\nu_x^+ = &\frac{\sqrt{1-\rho^2}}{2\sqrt{\pi}} \int_{-\infty}^{x/\sqrt{2(1-\rho^2)}} du\;\bigg\{e^{-(1-\rho^2)u^2}\, \\ 
&\times\left[1-\mathrm{erf}\left(x/\sqrt{2(1-\rho^2)}-\rho u \right)\right]\bigg\},
\end{align}
where
\begin{equation}
\mathrm{erf}(x) = \frac{2}{\sqrt{\pi}}\int_0^x dt\;e^{-t^2},
\end{equation}
is the error function. It is important to note that the level-crossing result for discrete-time stationary Gaussian processes in Eq.~(\ref{eq: 1d nu for Gaussian processes}) obviously differs from the result for continuous-time stationary Gaussian processes in Eq.~(\ref{eq: LC for stationary Gaussian SP}). In the special case of no correlation, $\rho=0$, i.e. when $\{X\}$ is Gaussian white noise, Eq.~(\ref{eq: 1d nu for Gaussian processes}) could be simply written as
\begin{equation}
\label{eq: nu for Gaussian white noise}
\nu_x^+ = \frac{1}{4} \left[ 1-\mathrm{erf}^2\left( x/\sqrt{2} \right) \right].
\end{equation}
The average frequency $\nu_x^+$ in Eq.~(\ref{eq: 1d nu for Gaussian processes}) for several values of the correlation $\rho$ is depicted in Fig.~\ref{fig: one-dimensional level-crossing frequency}. The curve for $\rho=0$ in Fig.~\ref{fig: one-dimensional level-crossing frequency} is for the case of Gaussian white noise. As seen in Fig.~\ref{fig: one-dimensional level-crossing frequency}, the average number of upcrossings of the level $x$ increases by decreasing the correlation $\rho$. This observation could be justified by the fact that a negatively (positively) correlated process, $\rho<0$ ($\rho>0$) fluctuates more (less) than a white noise ($\rho=0$) which in turn leads to increasing (decreasing) the chance of upcrossing. It is noteworthy to state that the analytic description [Eq.~(\ref{eq: 1d nu for Gaussian processes})] provided in this section confirms the numerical results obtained for fractional Gaussian noises using the level-crossing method, see Ref.~\cite{Vahab}. 

\section{Level crossing for discrete-time vector stochastic processes}

The level-crossing method proposed and implemented in studies prior to this work were basically able to analyze just a single continuous-time stochastic process or, at best, to analyze multiple processes in an individual manner \cite{JMFRM,BSJ}. These are not options in this work, due to the fact that processes are not always individual or independent of each other. To comply with the aims of the present study where the coupling of the stochastic processes is not neglected, a vector stochastic process needs to be considered~\cite{comment2}, and hence a new method needs to be provided. As such, we develop the level-crossing technique so we can call it the generalized level-crossing method.

Mathematically speaking, a vector stochastic process is considered a $d$-dimensional path represented by $\{\mathbf{X}\}\equiv\{\mathbf{X}(t_1),\mathbf{X}(t_2),\dots,\mathbf{X}(t_n)\}$; where we have $\mathbf{X}(t_i)=(X_1(t_i),X_2(t_i),\dots,X_d(t_i))$. The index $d\,(\ge2)$  indicates the number of stochastic processes being handled at the same time instance. To state clearer, when dealing with two simultaneous stochastic processes, e.g. the price return fluctuations of oil and gold, the parameter $d$ would be equal to 2, which is due to the fact that two markets are being considered. We propose our apparatus for extracting statistical information on the path of a vector stochastic process $\{\mathbf{X}\}$. This is what we call the generalized level crossing for a vector stochastic process in the $d$ dimension. This method is founded on the combination of two concepts; radial and angular level crossings.

\subsection{Radial level crossing}

Consider a $d$-dimensional vector stochastic process, $\{\mathbf{X}\}$, and a $d$-dimensional sphere of radius $R$ centered at the origin. In typical level-crossing methods, a level is crossed by a stochastic process. In the model studied here, the vector stochastic process ($\{\mathbf{X}\}$) passes through the surface of a $d$-dimensional sphere, see  Fig.~\ref{fig: d-dimensional level-crossing}(a) for a three dimensional illustration. Now we define the radial level crossing for a vector stochastic process in a similar manner to the definition of the level crossing for a stochastic process. In the time period from $t_1$ to $t_n$ the desired path $\{\mathbf{X}\}$ outcrosses our sphere $N_n^+(R)$ times. As in the one-dimensional case, an outcrossing of a $d$-dimensional sphere of radius $R$ at time $t_i$ is when the conditions $X(t_{i-1})< R$ and $X(t_{i+1})> R$ are complied. Note that $X$ is the distance of the point $\mathbf{X}$ from the origin. Since $\{\mathbf{X}\}$ is a stochastic path, $N_n^+(R)$ would posses a random behavior. Therefore its ensemble average, which is denoted by  $\langle N_n^+(R) \rangle$, would be our desired parameter. The average number of outcrossings depends on two facts:  first, on the radial distribution of the points on the path $\{\mathbf{X}\}$ and, second, on the radial correlation of $\{\mathbf{X}\}$. In the case of a stationary vector stochastic process, $\langle N_n^+(R) \rangle$ would become proportional to $n$ with a proportionality constant $\nu_R^+$~\cite{comment3}. Since, $\nu_R^+$ is the average frequency of the surface (with radius $R$) outcrossings, its inverse represented by $\tau_{R}=1/\nu_{R}^+$ gives the radial waiting time between two successive outcrossings.

Quite similar to the one-dimensional case of the previous section, the average frequency $\nu_R^+$ in the $d$ dimension for a stationary vector stochastic process $\{\mathbf{X}\}$ is given by
\begin{equation}
\label{eq: d-dimensional radial nu}
\nu_R^+ = \int_{x_1<R} d^dx_1 \int_{x_2>R} d^dx_2 ~ p_{\mathbf{X}}(\boldsymbol{x}_1,\boldsymbol{x}_2),
\end{equation}
where $p_{\mathbf{X}}(\boldsymbol{x}_1,\boldsymbol{x}_2)$ is the two-point joint probability density of the vector stochastic process $\{\mathbf{X}\}$ for the two successive points $\boldsymbol{x}_1$ and $\boldsymbol{x}_2$. So, the right-hand side of Eq.~(\ref{eq: d-dimensional radial nu}) is the occurrence probability for an outcrossing of the surface of the $d$-dimensional sphere with radius $R$ by two successive points of $\{\mathbf{X}\}$. Due to the importance of Gaussian processes we consider $\{\mathbf{X}\}$ to be a stationary vector stochastic process with standardized Gaussian distribution. In this case, we have
\begin{equation}
\label{eq: multivariate Gaussian pdf}
p_{\mathbf{X}}(\boldsymbol{x}_1,\boldsymbol{x}_2) = \frac{1}{(2\pi)^{d}}\frac{1}{\sqrt{det K}} \exp{\left\{-\frac{1}{2}P^T K^{-1} P\right\}},
\end{equation}
where $P = (\boldsymbol{x}_1,\boldsymbol{x}_2)^T$ is a column vector containing coordinates of the two successive points, the superscript ``T'' indicates the transpose operation, and $K = \langle PP^T\rangle$ is the covariance matrix.  All information about correlations between the two points $\boldsymbol{x}_1$ and $\boldsymbol{x}_2$ is placed within the covariance matrix $K$. In spite of having $p_{\mathbf{X}}(\boldsymbol{x}_1,\boldsymbol{x}_2)$ for Gaussian vector stochastic processes, the double integral of Eq.~(\ref{eq: d-dimensional radial nu}) cannot be further simplified unless it is for the special case of no correlation, where we refer to it in the next section.

Although in this stage the radial level crossing is understood, valuable information from the vector stochastic process still could not be extracted, since the radial level crossing tells us nothing about the angular behavior of the vector stochastic process. This brings need for the presentation of the angular level crossing.

\subsection{Angular level crossing}

\begin{figure}[t!]
\centering
\includegraphics[width=1\linewidth]{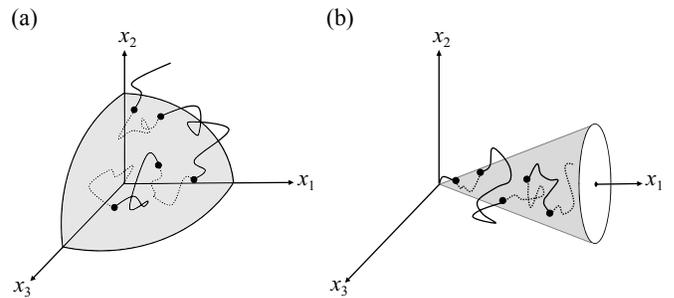}
\caption{A schematic illustration of the radial (a) and angular (b) level crossings. Panel (a): A cut of a whole sphere limited to the positive axis of the Cartesian coordinates. The dashed-solid line sketches a path taken by a vector stochastic process named $\{\mathbf{X}\}$. Panel (b): A cone where its axis overlaps with one of the Cartesian axes. Note that for both panels the black points are created when the path crosses the surface of the shapes, and the dashed and solid parts of the line indicate whether its inside or outside the shapes. }
\label{fig: d-dimensional level-crossing}
\end{figure}

Consider one of the Cartesian axis $x_i$ where, at the same time, is the axis of a cone, see Fig.~\ref{fig: d-dimensional level-crossing}(b) for a three dimensional illustration. The cone apex is located at the origin of the coordinate system. The cone is specified by its apex angle, $\alpha$, that takes values between $0$ to $\pi$. Now the average number of outcrossings that path $\{\mathbf{X}\}$ experiences through the side surface of the cone represented by $\langle N_n^+(\alpha) \rangle$, is what we are going to count. Note that the average number of outcrossings depends on two facts. Firstly, on the angular distribution of the points on the path $\{\mathbf{X}\}$, and secondly, on the angular correlation of $\{\mathbf{X}\}$. In case of a stationary vector stochastic process,  $\langle N_n^+(\alpha) \rangle$, would become proportional to $n$ with a proportionality constant $\nu_{\alpha}^+$. Since, $\nu_{\alpha}^+$ is the average frequency of the outcrossings through the side surface of the cone, its inverse represented by $\tau_{\alpha}=1/\nu_{\alpha}^+$ gives us the angular waiting time. The angular waiting time $\tau_{\alpha}$ is the the average time between two consecutive side- surface outcrossings by $\{\mathbf{X}\}$.

As in the radial case, the average frequency $\nu_{\alpha}^+$ for a stationary vector stochastic process $\{\mathbf{X}\}$ is given by
\begin{equation}
\label{eq: d-dimensional angular nu}
\nu_{\alpha}^+ = \int_{\vartheta_1<\alpha/2} d^dx_1 \int_{\vartheta_2>\alpha/2} d^dx_2 ~ p_{\mathbf{X}}(\boldsymbol{x}_1,\boldsymbol{x}_2),
\end{equation}
where $p_{\mathbf{X}}(\boldsymbol{x}_1,\boldsymbol{x}_2)$ is the two-point joint probability density of the vector stochastic process $\{\mathbf{X}\}$. Angles $\vartheta_1$ and $\vartheta_2$ in Eq.~(\ref{eq: d-dimensional angular nu}) are made by the vectors $\boldsymbol{x}_1$ and $\boldsymbol{x}_2$ with the $x_i$ axis ,respectively.  When $\{\mathbf{X}\}$ represents a stationary Gaussian vector stochastic process, the same explanations expressed for the radial level crossing applies.

\section{The creation of a criterion}

However, we are still not there yet. The reason is that when implementing Eqs. (\ref{eq: d-dimensional radial nu}) and (\ref{eq: d-dimensional angular nu}) for radial and angular level crossings, the results are not conclusive by themselves. This is due to the fact that the results should first be valued. In other words they need to be compared with some sort of a criterion to provide a basis for the most suitable and applicable conclusions. Usually the best criterion that would work as a measure must not be biased. Therefore, in the context of the present study, the criterion is selected to be an uncorrelated Gaussian process. The intention is to obtain an analytic solution for the frequencies $\nu_{R}^+$ and $\nu_{\alpha}^+$ of a vector stochastic process in $d$ dimensions whose components are independent Gaussian white noises.

To obtain the criterion, consider a vector stochastic process consisting of $d$ independent Gaussian white noises represented by $\{\mathbf{X}_{wn}\}$. The joint probability distribution for two successive points $\boldsymbol{x}_1$ and $\boldsymbol{x}_2$ of the path $\{\mathbf{X}_{wn}\}$ is given by Eq.~(\ref{eq: multivariate Gaussian pdf}) in which the covariance matrix $K$ is equal to the identity matrix $I$. The joint probability distribution $p_{\mathbf{X}}(\boldsymbol{x}_1,\boldsymbol{x}_2)$ is then reduced to
\begin{equation}
\label{eq: joint pdf for Gaussian WN}
p_{\mathbf{X}}(\boldsymbol{x}_1,\boldsymbol{x}_2) = p(\boldsymbol{x}_1)p(\boldsymbol{x}_2),
\end{equation}
with  
\begin{equation}
\label{eq: Gaussian WN pdf}
p(\boldsymbol{x}) = \frac{1}{(\sqrt{2\pi})^d} \exp{\left(-\frac{1}{2}\boldsymbol{x}.\boldsymbol{x}^T \right)}.
\end{equation}
Equation~(\ref{eq: joint pdf for Gaussian WN}) means that there is no correlation between the two points leaving them independent of each other. Using this equation, the double integral of Eq.~(\ref{eq: d-dimensional radial nu}) is split into the product of two univariate integrals. So, for the vector stochastic process $\{\mathbf{X}_{wn}\}$, the radial frequency is given by
\begin{equation}
\label{eq: nu_R white noise}
\nu_R^+ = P_{in} P_{out},
\end{equation}
where
\begin{equation}
\label{eq: Radial P_in}
P_{in}=\int_{x_1<R} d^d{x}_1\;p(\boldsymbol{x}_1), 
\end{equation}
and
\begin{equation}
\label{eq: Radial P_out}
P_{out}=\int_{x_2>R} d^d{x}_2\;p(\boldsymbol{x}_2).
\end{equation}
As seen from the domain of integration in Eqs.~(\ref{eq: Radial P_in}) and (\ref{eq: Radial P_out}), $P_{in}$ represents the probability of the first point to be inside the sphere of radius $R$ centered at the origin, and $P_{out}$ is the probability of the second point to be outside the very sphere. Since every point on the path is either inside or outside the sphere, the summation of the probabilities $P_{in}$ and $P_{out}$ is equal to unity. Using Eq.~(\ref{eq: Gaussian WN pdf}) the probability ($P_{in}$) is obtained as follows:
\begin{equation}
\label{eq: Exact form of radial P_in for white noise}
P_{in} = \frac{2^{(1-d/2)}}{(d/2-1)!}\,\int_0^R dx\;x^{d-1} e^{-x^2/2}.
\end{equation}
The integration of Eq.~(\ref{eq: Exact form of radial P_in for white noise}) can be calculated using the method of integration by parts. This enables an analytical solution for the positive frequencies, where by substituting the expressions for $P_{in}$ and $P_{out}=1-P_{in}$ from Eq.~(\ref{eq: Exact form of radial P_in for white noise}) into Eq.~(\ref{eq: nu_R white noise}), $\nu_R^+$ is obtained.

Due to the isotropic characteristic of the path $\{\mathbf{X}_{wn}\}$, there is no preferred direction. Therefore, we could obtain $\nu^+_{\alpha}$ in Eq.~(\ref{eq: d-dimensional angular nu}) for an arbitrary axis, $x_i$, and say that this result would be the same for all other directions. The angle between a typical point $\boldsymbol{x}$ on the curve $\{\mathbf{X}_{wn}\}$ and the $x_i$ axis is denoted by $\vartheta$. Since the two successive points $\boldsymbol{x}_1$ and $\boldsymbol{x}_2$ on the path $\{\mathbf{X}_{wn}\}$ are independent random vectors, the angles $\vartheta_1$ and $\vartheta_2$ are independent random variables. This independence enables us to write the similar equation as Eq.~(\ref{eq: nu_R white noise}) for the case of $\nu^+_{\alpha}$. So the angular frequency is given by
\begin{equation}
\label{eq: nu_alpha white noise}
\nu^+_{\alpha} = P_{in} P_{out},
\end{equation}
with
\begin{equation}
\label{eq: Angular P_in}
P_{in}=\int_{\vartheta_1<\alpha/2} d^d{x}_1\;p(\boldsymbol{x}_1), 
\end{equation}
and
\begin{equation}
\label{eq: Angular P_out}
P_{out}=\int_{\vartheta_2>\alpha/2} d^d{x}_2\;p(\boldsymbol{x}_2).
\end{equation}
Note that, in this case, $P_{in}$ ($P_{out}$) is the probability that a point resides inside (outside) a cone with the apex on the origin, the axis overlaps with the $x_i$ axis, and the apex angle is denoted by $\alpha$. By substituting the probability density function of Eq.~(\ref{eq: Gaussian WN pdf}) into Eq.~(\ref{eq: Angular P_in}) and integrating over the solid angle of the cone, the probability $P_{in}$ is obtained as follows:
\begin{equation}
\label{eq: Exact form of angular P_in for white noise 1}
P_{in} =\left\lbrace
\begin{array}{ll}
\alpha/2\pi & d = 2 \\ \\
\sin^2(\alpha/4) & d = 3 ,
\end{array}
\right.
\end{equation}
for the lower dimensions $d=2,3$, and 
\begin{equation}
\label{eq: Exact form of angular P_in for white noise 2}
P_{in}=\sqrt{\frac{2}{\pi}}\; \Gamma\!\left( \frac{1+d}{2} \right) \prod_{k=1}^{d-3} \frac{\Gamma\!\left( \frac{1+k}{2} \right)}{\Gamma\!\left( 1+\frac{k}{2} \right)} \int_0^{\alpha/2} d\vartheta \;(\sin{\vartheta})^{d-2},
\end{equation}
for the higher dimensions $d>3$, where $\Gamma(x)$ is the gamma function. This enable us to have an analytical solution for the positive frequency, $\nu_{\alpha}^+=P_{in}P_{out}$, which is obtained by the expressions for $P_{in}$ and the equality $P_{out}=1-P_{in}$. Note that for extracting Eq.~(\ref{eq: Exact form of angular P_in for white noise 2}) we used the hyperspherical coordinate system which is the generalization of the spherical coordinate system to the dimensions higher than $3$.

The interest here is to browse the collective behavior of a multiple process by crossing a specific level. In other words, we show how coupling is featured in the context of the present study. Toward this end, we investigate the generalized level-crossing method in two dimensions. The positive frequencies for the criteria in two dimensions is obtained by substituting the expressions for $P_{in}$ and $P_{out} = 1-P_{in}$ from Eqs.~(\ref{eq: Exact form of radial P_in for white noise}) and (\ref{eq: Exact form of angular P_in for white noise 1}) into Eqs.~(\ref{eq: nu_R white noise}) and (\ref{eq: nu_alpha white noise}), 
\begin{equation}
\label{eq: radial frequency for uncoupled case}
\nu_0^+(R) = e^{-R^2/2}\left(1-e^{-R^2/2}\right),
\end{equation}
and
\begin{equation}
\label{eq: angular frequency for uncoupled case}
\nu_0^+(\alpha) = \frac{\alpha}{2\pi}\left(1- \frac{\alpha}{2\pi}\right),
\end{equation}
where the zero index is used to emphasize that these frequencies are for the uncoupled case and distinguish them from the coupled case. Now, consider a vector stochastic process $\{\mathbf{X}\}=\{(x_1,y_1),(x_2,y_2),\dots\}$ consisting of two standardized Gaussian white noises $\{x_1,x_2,\dots\}$ and $\{y_1,y_2,\dots\}$ with a Gaussian coupling as
\begin{equation}
\label{eq: coupling between two white noises}
C(n)=\langle x_i y_{i+n}\rangle = A \exp\left(-\frac{n^2}{\xi^2}\right),
\end{equation}
where $\langle.\rangle$ denotes the ensemble average, ``$A$'' represents the amplitude of coupling, and ``$\xi$'' is the correlation length. In this case, there exists no analytical expression  for the average frequencies $\nu_R^+$ and $\nu_\alpha^+$. Therefore they could only be computed by the numerical integrations of Eqs.(\ref{eq: d-dimensional radial nu}) and (\ref{eq: d-dimensional angular nu}). Note that Eq.~(\ref{eq: coupling between two white noises}) determines the elements of the covariance matrix, $K$, in Eq.~(\ref{eq: multivariate Gaussian pdf}). 

\begin{figure}[b!]
\centering
\includegraphics[width=0.97\linewidth]{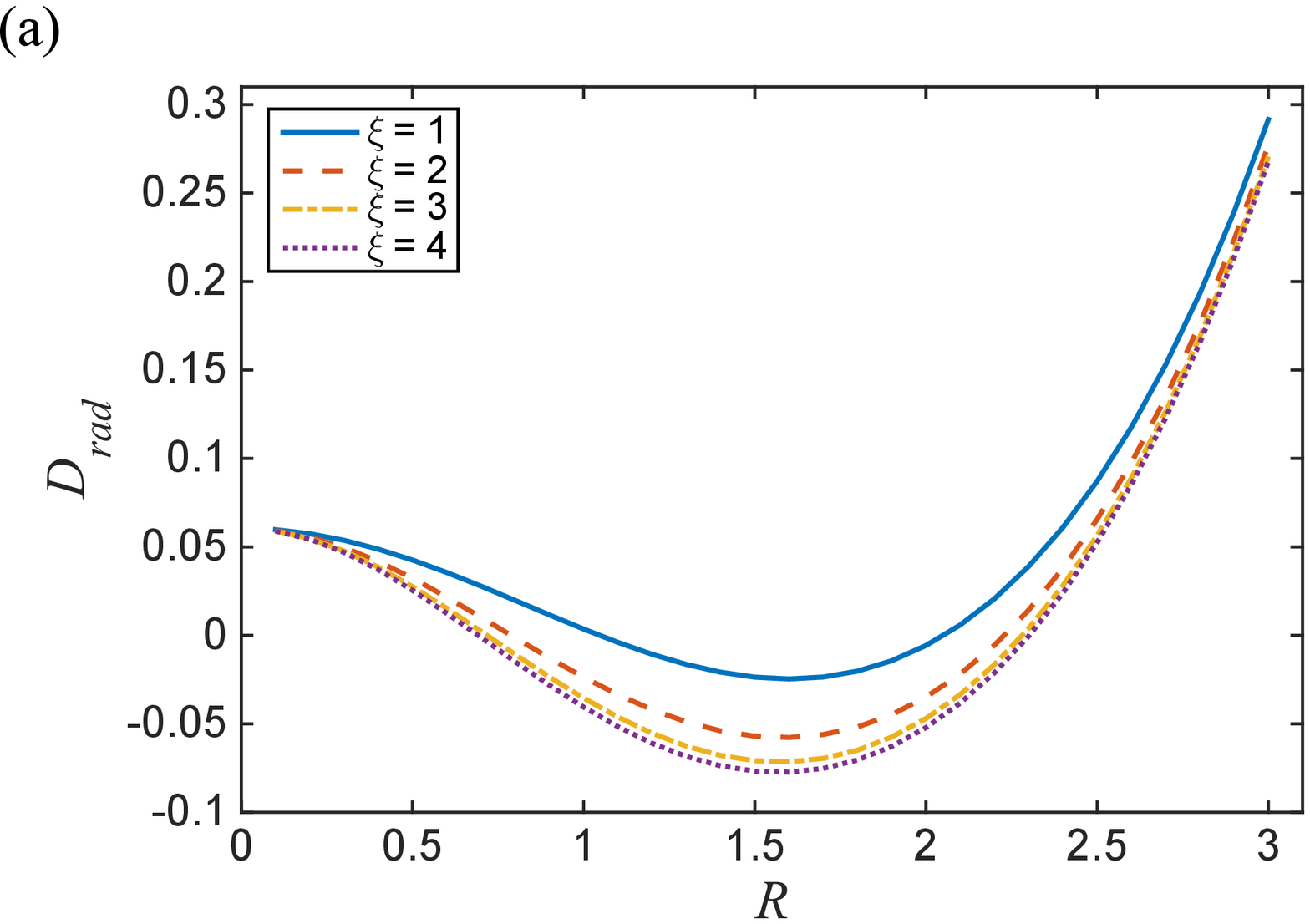}
\vspace{3mm}

\includegraphics[width=0.97\linewidth]{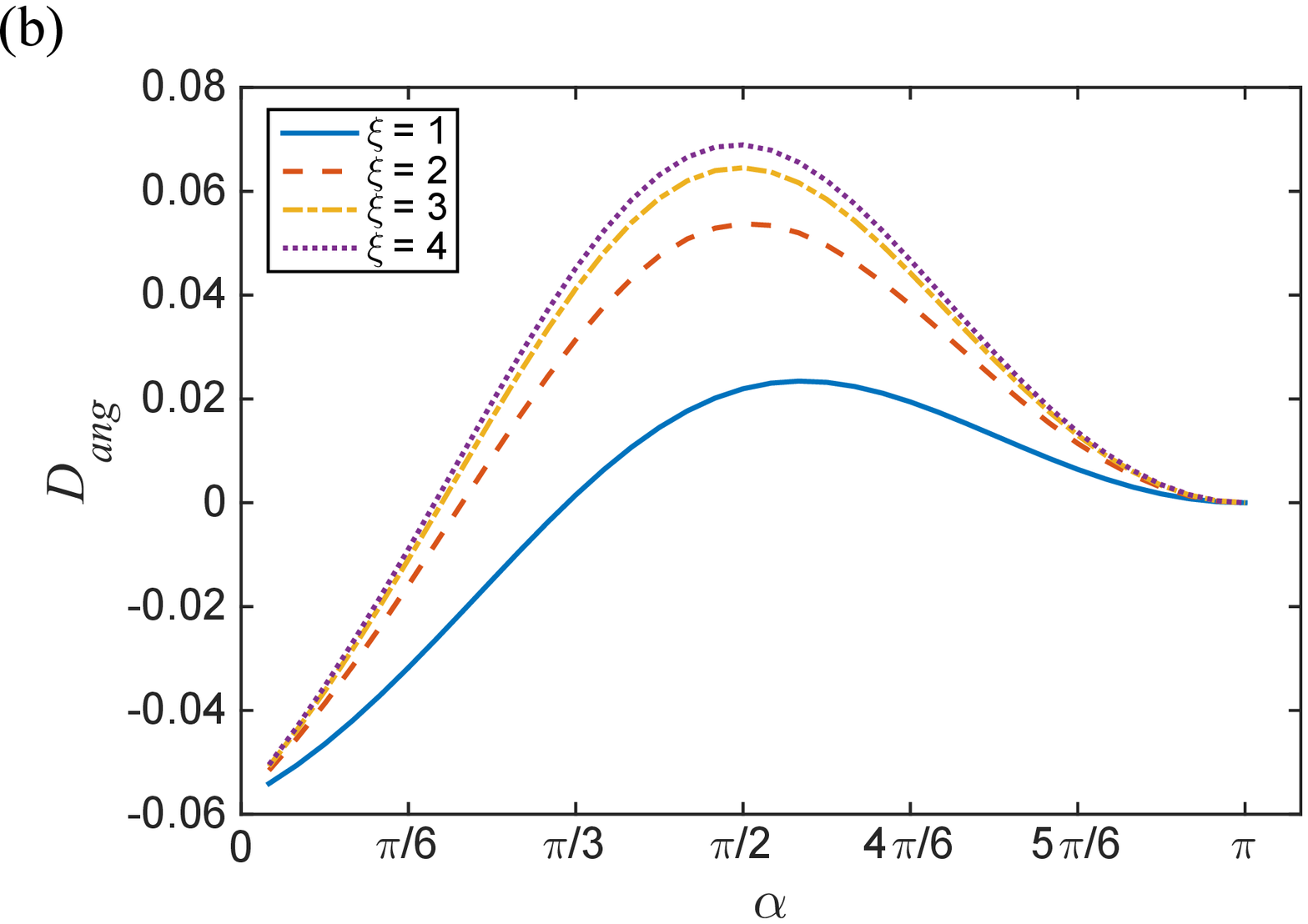}
\caption{Sensitivity of the two dimensional level-crossing analysis on coupling with short correlation lengths. Panels (a) and (b) are the result of comparison between two uncoupled white noises and two coupled white noises, respectively, in the radial and angular level-crossing regimes.}
\label{fig: coupling effect}
\end{figure}

In order to show the deviation between the frequencies $(\nu_R^+,\nu_\alpha^+)$ of coupled white noises and $(\nu_0^+(R),\nu_0^+(\alpha))$ of uncoupled white noises, we introduce the following ratios
\begin{equation}
D_{rad}=\frac{\nu_R^+ -\nu_0^+(R)}{\nu_0^+(R)},
\end{equation}
and
\begin{equation}
D_{ang}=\frac{\nu_\alpha^+ -\nu_0^+(\alpha)}{\nu_0^+(\alpha)}.
\end{equation}
Figure \ref{fig: coupling effect} shows these deviation ratios for radial [Fig.~\ref{fig: coupling effect}(a)] and angular [Fig.~\ref{fig: coupling effect}(b)] crossings for the Gaussian coupling in Eq.~(\ref{eq: coupling between two white noises}) with the amplitude $A=1/3$ and correlation lengths $\xi=1,2,3,4$. Notice that the deviation is directly proportional to the correlation length, where a higher correlation length gives a curve with a bigger deviation. This is true for both radial and angular level crossings. Another important conclusion made from Figs. \ref{fig: coupling effect}(a) and \ref{fig: coupling effect}(b) comes from the size of the correlation length $\xi$. The fact of the matter is that this itself is a standing point for the promise of the present study. In other words, although the correlation length is very small, there exists a pronounced deviation between the coupled and uncoupled cases.

\section{Conclusions}

All processes in nature, although they may seem to be continuous, are actually not. As a matter of fact, all processes are discrete, time wise. In other words, a seemingly continuous process is a look from far at that process. Now by looking closer and closer, its discreteness becomes observable. Toward that end, we extended the level-crossing method to the realm of discrete-time stochastic processes, which enabled filling two of the important gaps in this method. First, we obtained an analytical expression for the level crossing of a discrete-time stationary Gaussian process, see Eq.~(\ref{eq: 1d nu for Gaussian processes}). The generality of the expression is due to the fact that it is derived for a Gaussian process with an arbitrary correlation. Second, we developed the level-crossing method to enable simultaneous analysis of several discrete-time processes. The reason for going this way is due to the fact that processes are not exactly independent. Our analytic modeling and hence solutions contribute towards better understanding this statement. The generalized level-crossing method consists of two working concepts, namely radial and angular level crossing. These new concepts enable us to study the coupling between processes. In order to pronounce the efficiency of the coupling between the components of a multiple process, we introduced the state-of-the-art criterion which is the benchmark of multiple processes with no coupling. We derived analytic results for the radial and angular level-crossing regarding this criterion, see Eqs. (\ref{eq: nu_R white noise}) and (\ref{eq: nu_alpha white noise}). In order to evaluate the criterion, the generalized level-crossing has been studied in two dimensions. The results show the sensitivity of the radial and angular solutions to slight couplings with short correlation lengths.

\bibliography{generelized_level_crossing}

\begin{thebibliography}{35}
\bibitem [{com({\natexlab{a}})}]{comment1}%
  \BibitemOpen
  \href@noop {} {}\bibinfo {note} {The standardized Gaussian process is a
  Gaussian process with zero mean and unit variance.}\BibitemShut {Stop}%
\bibitem [{com({\natexlab{b}})}]{comment2}%
  \BibitemOpen
  \href@noop {} {}\bibinfo {note} {Vector stochastic processes are the proper
  mathematical framework for studying multiple stochastic
  processes.}\BibitemShut {Stop}%
\bibitem [{com({\natexlab{c}})}]{comment3}%
  \BibitemOpen
  \href@noop {} {}\bibinfo {note} {Note that a vector stochastic process is
  called stationary when all its components are stationary.}\BibitemShut
  {Stop}%
\end{thebibliography}

\end{document}